\documentclass[twocolumn,preprintnumbers,amsmath,amssymb]{revtex4}

\usepackage{graphicx}
\usepackage{dcolumn}
\usepackage{bm}

\DeclareMathOperator{\coul}{coul}
\DeclareMathOperator{\erf}{erf}
\DeclareMathOperator{\erfgau}{erfgau}
\newcommand{\Psimu}{\ensuremath{\Psi^{\mu}}}
\newcommand{\bra}[1]{\ensuremath{\langle #1 \vert}}
\newcommand{\ket}[1]{\ensuremath{\vert #1  \rangle}}

\renewcommand{\b}[1]{\ensuremath{\mathbf{#1}}}

\begin{document}

\title{Short-range exchange-correlation energy of a uniform electron gas with modified electron-electron interaction}

\author{Julien Toulouse}
\author{Andreas Savin}
 \email{savin@lct.jussieu.fr}
\affiliation{
Laboratoire de Chimie Th\'eorique, CNRS et Universit\'e Pierre et Marie Curie,\\
4 place Jussieu, 75252 Paris, France
}

\author{Heinz-J\"{u}rgen Flad}
\affiliation{
 Max-Planck-Institut f\"{u}r Mathematik in den Naturwissenschaften,\\
 Inselstr. 22-26, D-04103 Leipzig, Germany \\
}

\date{\today}

\begin{abstract}
We calculate the short-range exchange-correlation energy of the uniform electron gas with two modified electron-electron interactions. While the short-range exchange functionals are calculated analytically, Coupled-Cluster and Fermi-hypernetted chain calculations are carried out for the correlation energy and the results are fitted to an analytical parametrization. These data enable to construct the local density approximation corresponding to these modified interactions.
\end{abstract}

\maketitle

\section{Introduction}
\label{sec:Intro}

In electronic structure calculations using density functional theory (DFT)~\cite{HohKoh-PR-64} in the Kohn-Sham (KS) scheme~\cite{KohSha-PR-65}, the central quantity that needs to be approximated is the exchange-correlation energy functional. The vast majority of approximations for this functional originates from the local density approximation (LDA)~\cite{KohSha-PR-65} consisting in locally transferring the exchange-correlation energy of the uniform electron gas to the inhomogeneous system of interest. Actually, it is been realized for long that the LDA can describe accurately short-range correlation effects but is inadequate for long-range correlation effects (see, e.g.,~\cite{LanPer-PRB-77}). This observation lead to the development of the first gradient corrected functionals~\cite{LanMeh-PRL-81,LanMeh-PRB-83,Per-PRL-85,PerWan-PRB-86,Per-PRB-86,Per-INC-91} with the basic idea that the long-range contribution to the exchange-correlation energy of the uniform electron gas must not be transferred to inhomogeneous systems.

Guided by the same idea, it has been proposed~\cite{StoSav-INC-85,Sav-INC-96,LeiStoWerSav-CPL-97,PolSavLeiSto-JCP-02,PolColLeiStoWerSav-IJQC-03,SavColPol-IJQC-03} to describe only the short-range electronic correlations effects by a density functional, and leaving the remaining long-range correlations effects to a more appropriate method like Configuration Interaction. Concretely, the method is based on a decomposition of the true Coulomb electron-electron interaction as
\begin{equation}
\label{}
\frac{1}{r} = v_{ee}^{\mu}(r) + \bar{v}_{ee}^{\mu}(r),
\end{equation}
where $v_{ee}^{\mu}(r)$ is a long-range interaction and $\bar{v}_{ee}^{\mu}(r)$ is the complement short-range interaction. This separation is controlled by the parameter $\mu$. In previous studies~\cite{Sav-INC-96,LeiStoWerSav-CPL-97,PolSavLeiSto-JCP-02}, the error function has been used to define the long-range interaction
\begin{equation}
\label{}
v_{ee,\erf}^{\mu}(r)=\frac{\erf(\mu r)}{r},
\end{equation}
referred to as the \textit{erf} interaction. More recently~\cite{TouColSav-JJJ-XX}, we have used a sharper long-range/short-range separation with the \textit{erfgau} interaction
\begin{equation}
\label{}
v_{ee,\erfgau}^{\mu}(r)=\frac{\erf(\mu r)}{r} - \frac{2\mu}{\sqrt{\pi}} e^{-\frac{1}{3}\mu^2 r^2}.
\end{equation}
The method then consists in finding the ground-state multi-determinantal wave function $\Psimu$ of a fictitious system containing only the long-range part of the electron-electron interaction $\hat{V}_{ee}^{\mu}=\sum_{i<j}v_{ee}^{\mu}(r_{ij})$ and having the same density $n$ than the physical system.
The total ground-state electronic energy of a physical system is then given by
\begin{equation}
E =  \bra{\Psimu} \hat{T}+\hat{V}_{ee}^{\mu}+\hat{V}_{ne} \ket{\Psimu} + \bar{U}^{\mu}[n] + \bar{E}^{\mu}_{xc}[n],
\label{E}
\end{equation}
where $\hat{T}$ is the kinetic energy operator, $\hat{V}_{ne}$ is the nuclei-electron interaction, $\bar{U}^{\mu}$ is the short-range Hartree energy and $\bar{E}^{\mu}_{xc}$ is the short-range exchange-correlation functional defined as the difference between the standard KS exchange-correlation energy $E_{xc}$ and the long-range exchange-correlation energy $E^{\mu}_{xc}$ associated to the interaction $v_{ee}^{\mu}$
\begin{equation}
\label{}
\bar{E}^{\mu}_{xc} = E_{xc} - E^{\mu}_{xc}.
\end{equation}
Eq.~(\ref{E}) provides an exact decomposition of the total energy into a long-range component written in a wave function formalism and a remaining short-range component expressed as a density functional. In particular, there is no double counting of correlation effects. The only unknown quantity in this approach is the short-range exchange-correlation functional $\bar{E}^{\mu}_{xc}$ which is not the usual exchange-correlation functional of the KS scheme $E_{xc}$.

For a reasonable long-range/short-range separation ($\mu$ not too small), $\bar{E}^{\mu}_{xc}$ essentially describes short-range interactions, and it is therefore expected to be well approximated by the LDA corresponding to the modified interaction
\begin{equation}
\bar{E}^{\mu}_{xc}[n]= \int n(\b{r}) \bar{\varepsilon}^{\mu}_{xc}(n(\b{r})) d\b{r}.
\label{LDA}
\end{equation}
In Eq.~(\ref{LDA}), $\bar{\varepsilon}^{\mu}_{xc}$ is the short-range exchange-correlation energy per particle obtained by difference from the exchange-correlation energies per particle of the uniform electron gas with the standard Coulomb $\varepsilon_{xc}$ and with the \textit{erf} or \textit{erfgau} interaction $\varepsilon^{\mu}_{xc}$
\begin{equation}
\label{}
\bar{\varepsilon}^{\mu}_{xc}(n) = \varepsilon_{xc}(n) - \varepsilon^{\mu}_{xc}(n).
\end{equation}

As for the original LDA in the KS scheme with the Coulomb interaction, knowledge of $\bar{\varepsilon}^{\mu}_{xc}$ is crucial to apply the LDA to the short-range exchange-correlation functional. In this paper, we give the expressions of this short-range exchange-correlation energy per particle of the uniform electron gas with the \textit{erf} and \textit{erfgau} modified interactions. While the exchange part can be calculated analytically, the correlation are derived from Coupled-Cluster and from Fermi-hypernetted chain calculations.

Atomic units will be used throughout this work.

\section{Short-range exchange energy}
\begin{figure}
\includegraphics[scale=0.7]{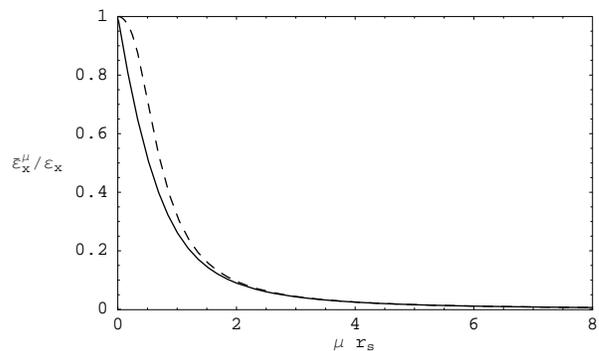}
\caption{Ratio of the \textit{erf} (solid line) or \textit{erfgau} (dashed line) short-range exchange energy per particle to the exchange energy per particle with Coulomb interaction in the uniform electron gas $\bar{\varepsilon}^{\mu}_{x}(r_s)/\varepsilon_{x}(r_s)$ with respect to $\mu r_s$. In order to compare the two interactions, a scale factor has been applied on the interaction parameter of the \textit{erfgau} interaction: $\mu \to 3.375 \mu$.
}
\label{fig:exsrerferfgau}
\end{figure}

The short-range exchange energies per particle $\bar{\varepsilon}^{\mu}_{x}(r_s)$ of the uniform electron gas for the Wigner-Seitz radius $r_{s}=(3/(4\pi n))^{1/3}$ with the \textit{erf} and \textit{erfgau} interactions are calculated analytically (see Eq.~\ref{exsrerf} and~\ref{exsrerfgau} of Appendix~\ref{app:exchange}). The inverse of the interaction parameter, $1/\mu$, represents the range of the modified interaction and has to be compared with $r_s$ which is the characteristic length for exchange. Thus, the relevant variable for the exchange energy is actually $\mu r_s$. Fig.~\ref{fig:exsrerferfgau} shows the ratio of the short-range exchange energy per particle with the \textit{erf} and \textit{erfgau} interactions to the exchange energy per particle with the Coulomb interaction $\bar{\varepsilon}^{\mu}_{x}(r_s)/\varepsilon_{x}(r_s)$. In order to compare the two interactions, a scale factor is applied on the parameter of the \textit{erfgau} interaction $\mu \to (1+6\sqrt{3})^{1/2} \mu \approx 3.375 \mu$ so that the \textit{erf} and \textit{erfgau} exchange energies have the same asymptotic behavior for $\mu r_s \to \infty$ (see below).

It has been shown that for a finite system where the exchange contribution to the second-order density matrix $n_{2,x}(\b{r}_1,\b{r}_2)$ decays exponentially with $r_{12}$, the short-range exchange energy $\bar{E}_{x}^{\mu}$ can formally be expanded around $\mu=0$ into an odd series in $\mu$~\cite{TouColSav-JJJ-XX} 
\begin{eqnarray}
\bar{E}_{x}^{\mu} &=& E_{x} - \frac{1}{\sqrt{\pi}} \sum_{n=0}^{\infty} \frac{(-1)^n a_n}{n!} \mu^{2n+1} 
\nonumber\\
& & \times \iint n_{2,x}(\b{r}_1,\b{r}_2) r_{12}^{2n} d\b{r}_1 d\b{r}_2,
\label{Exmu0}
\end{eqnarray}
where $E_{x}$ is the usual KS exchange energy, $a_{n,\erf}=1/(2n+1)$ for the \textit{erf} interaction and $a_{n,\erfgau}=1/(2n+1) - 1/3^n$ ($\not= 0$ for $n \geq 2$) for the \textit{erfgau} interaction. Except for the term linear in $\mu$, the expansion for $\mu r_s \to 0$ of the \textit{erf} short-range energy per particle of the uniform electron gas does not exhibit the same behavior
\begin{eqnarray}
\bar{\varepsilon}^{\mu}_{x,\erf}(r_s) &\approx& \varepsilon_{x}(r_s) +\frac{1}{\sqrt{\pi}} \mu
- \left( \frac{3}{2\pi^4} \right)^{1/3} r_s \mu^2
+ \frac{2}{9\pi^2} r_s^3 \mu^4
\nonumber  \\
& & + \, \text{exponential terms }.
\end{eqnarray}
Similarly, the expansion corresponding to the \textit{erfgau} interaction is
\begin{eqnarray}
\bar{\varepsilon}^{\mu}_{x,\erfgau}(r_s) &\approx& \varepsilon_{x}(r_s) 
+  \frac{2\sqrt{3} -3}{(18\pi^4)^{1/3}} r_s \mu^2
+ \frac{2(9 -4\sqrt{3})}{81\pi^2} r_s^3 \mu^4 
\nonumber \\
& & + \, \text{exponential terms  }.
\label{epsxmu0erfgau}
\end{eqnarray}
These different behaviors of the short-range exchange energy in the uniform electron gas and in a finite system is consistent with the important LDA error arising at $\mu=0$, i.e. for the standard DFT within the Kohn-Sham scheme.

The short-range exchange energy of a finite system can also be formally expanded for $\mu \to \infty$ into the asymptotic series~\cite{TouSav-JJJ-XX}
\begin{equation}
\label{Exsrmuinf}
\bar{E}_{x}^{\mu} = 2\sqrt{\pi} \sum_{n=0}^{\infty} \frac{A_{2n}}{(2n)! (2n+2) \mu^{2n+2}}
\int n_{2,x}^{(2n)}(\b{r},\b{r})  d\b{r},
\end{equation}
where $n_{2,x}^{(2n)}(\b{r},\b{r})$ are the exchange contribution to the on-top second-order density matrix and its derivatives, $A_{n,\erf}=\Gamma(\frac{n+3}{2})$  for the \textit{erf} interaction and $A_{n,\erfgau}=\Gamma(\frac{n+3}{2}) - 3^{\frac{n+3}{2}} \Gamma(\frac{n+3}{2}) + 2 \times 3^{\frac{n+3}{2}} \Gamma(\frac{n+5}{2})$ for the \textit{erfgau} interaction. The asymptotic expansions of the short-range energies per particle of the uniform electron gas for large $\mu$ do have the same form
\begin{equation}
\label{}
\bar{\varepsilon}^{\mu}_{x,\erf}(r_s) \approx  -\frac{3}{16} \frac{1}{r_s^3 \mu^2}
+ \left( \frac{3\pi^2}{2} \right)^{1/3} \frac{27}{640} \frac{1}{r_s^5 \mu^4} + \cdots,
\end{equation}
\begin{eqnarray}
\bar{\varepsilon}^{\mu}_{x,\erfgau}(r_s) &\approx&   -\frac{3(1+6\sqrt{3})}{16} \frac{1}{r_s^3 \mu^2}
\nonumber \\
 & & + \left( \frac{3\pi^2}{2} \right)^{1/3} \frac{27(1+36\sqrt{3})}{640} \frac{1}{r_s^5 \mu^4} + \cdots.
\end{eqnarray}
Again, this is consistent with the quality of the local density approximation for large $\mu$~\cite{PolColLeiStoWerSav-IJQC-03,TouSav-JJJ-XX}.

\section{Short-range correlation energy}
\label{}

The long-range correlation energy per particle $\varepsilon_{c}^{\mu}({r_s})$ with the \textit{erfgau} interaction has been computed for several values of $r_s$ (from $r_s=0.2$ to $10$) and $\mu$ (from $\mu=0$ to $25$). Coupled-Cluster calculations with double excitations (CCD), according to a method introduced by Freeman~\cite{Fre-PRB-77}, have been performed (see Appendix~\ref{app:cc}), as well as Fermi-hypernetted-chain (FHNC) calculations (see Appendix~\ref{app:fhnc}). Data for the \textit{erf} interaction is already available~\cite{Sav-INC-96}.

Once the long-range correlation energy per particle $\varepsilon_{c}^{\mu}(r_s)$ is obtained, the short-range correlation energy per particle $\bar{\varepsilon}_{c}^{\mu }(r_s)$ is expressed as
\begin{equation}
\label{ecsr}
\bar{\varepsilon}_{c}^{\mu }(r_s) = \varepsilon_{c}(r_s) \left( 1- \frac{\varepsilon_{c}^{\mu}(r_s)}{\varepsilon_{c}^{\mu \to \infty}(r_s)} \right),
\end{equation}
where $\varepsilon_{c}(r_s)$ is the correlation energy per particle of the uniform electron gas with Coulomb interaction taken from the usual parametrization of Vosko, Wilk and Nusair (VWN)~\cite{VosWilNus-CJP-80}. According to Eq.~(\ref{ecsr}), $\bar{\varepsilon}_{c}^{\mu }(r_s)$ correctly reduces to the VWN value for $\mu=0$ and vanishes for $\mu \to \infty$.

\begin{figure}
\includegraphics[scale=0.6]{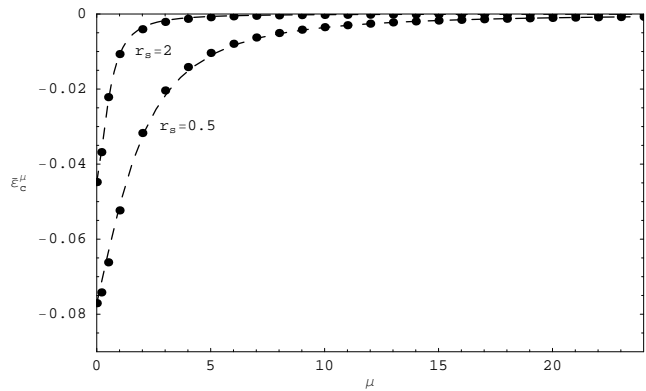}
\caption{Short-range correlation energy per particle (dots) of the uniform electron gas for the \textit{erf} interaction with respect to the interaction parameter $\mu$ for $r_s=0.5$ and $r_s=2$ computed with the CCD method. The analytical parametrization (Eq.~\ref{fit}) is represented by a dashed line.
}
\label{fig:ecsrerf}
\end{figure}

\begin{figure}
\includegraphics[scale=0.75]{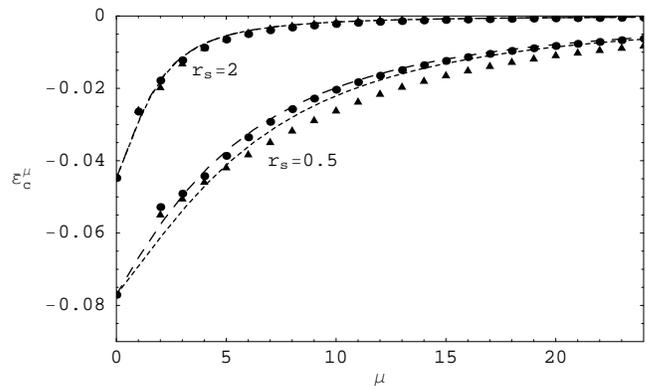}
\caption{Short-range correlation energy per particle of the uniform electron gas for the \textit{erfgau} interaction with respect to the interaction parameter $\mu$ for $r_s=0.5$ and $r_s=2$, computed with the CCD method (dots) and with the FHNC method (triangles). The analytical parametrizations (Eq.~\ref{fit}) using the CCD data and the FHNC data are represented by the long-dashed and short-dashed lines, respectively.
}
\label{fig:ecsrerfgau}
\end{figure}

The \textit{erf} and \textit{erfgau} short-range correlation energies per particle with respect to $\mu$ for $r_s=0.5$ and $r_s=2$ are plotted in Fig.~\ref{fig:ecsrerf} and~\ref{fig:ecsrerfgau}. For \textit{erfgau}, the differences between the results from the CCD and FHNC calculations are visible only for a high density ($r_s=0.5$). Surprisingly, both methods diverge with the \textit{erfgau} interaction when $\mu \sqrt{r_s} \lesssim 1$, explaining the absence of points between $\mu=0$ and $\mu \approx 0.7$ for $r_s=2$, or between $\mu=0$ and $\mu \approx 1.4$ for $r_s=0.5$ in Fig.~\ref{fig:ecsrerfgau}. We connect this behavior to the attractive character of the \textit{erfgau} interaction for small $\mu$ (see Appendix~\ref{app:divergence}). In practice, the lack of accuracy of the LDA correlation functional for small $\mu$ because of these missing points does not represent a serious problem since the LDA exchange functional produces large errors anyway in finite systems in this domain of $\mu$, as suggested by its incorrect expansion as $\mu \to 0$ (Eq.~\ref{epsxmu0erfgau}).

\begin{figure}
\includegraphics[scale=0.6]{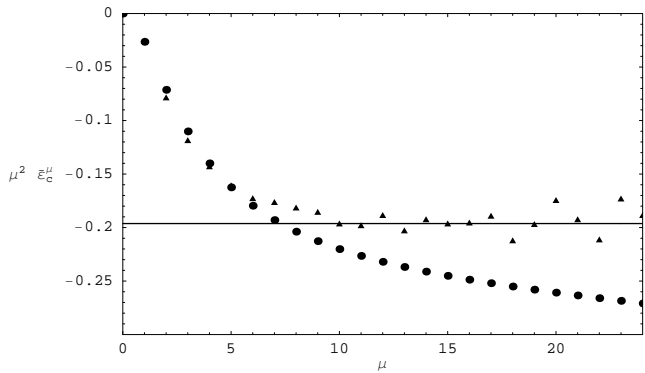}
\caption{ $\mu^2 \bar{\varepsilon}^{\mu}_{c}(r_s)$ computed from the CCD method (dots) and from the FHNC method (triangles) for the \textit{erfgau} interaction and $r_s=2$. The horizontal line is the exact limit for $\mu \to \infty$.
}
\label{fig:limitmuinf}
\end{figure}

It has been shown~\cite{PolColLeiStoWerSav-IJQC-03,TouSav-JJJ-XX} that the leading term in the expansion of the short-range correlation energy for large $\mu$ is
\begin{equation}
\label{}
\bar{E}_c^{\mu} \approx \frac{C \pi}{2 \mu^2} \int n_{2,c}(\b{r},\b{r}) d\b{r} + \cdots,
\end{equation}
where $n_{2,c}(\b{r},\b{r})$ is the on-top correlation pair density for the full Coulomb interaction, $C=1$ for the \textit{erf} interaction and $C=(1+6\sqrt{3})$ for the \textit{erfgau} interaction. For the uniform electron gas, $n_{2,c}(\b{r},\b{r})$ can be expressed in term of the on-top pair-distribution function $g_0(r_s)$ so that the short-range correlation energy per particle has the following exact behavior for $\mu \to \infty$
\begin{equation}
\label{}
\bar{\varepsilon}_c^{\mu}(r_s) \approx \frac{3 C}{8 \mu^2 r_s^3} \left( g_0(r_s) - \frac{1}{2} \right) + \cdots.
\end{equation}
An estimation of $g_0(r_s)$ which includes the correct limits for $r_s \to \infty$ and $r_s \to 0$ was given by Burke, Perdew and Ernzerhof~\cite{BurPerErn-JCP-98}
\begin{equation}
\label{}
g_0(r_s)=D \left( (\gamma + r_s )^{3/2} +\beta \right) e^{-A \sqrt{\gamma+r_s}},
\end{equation}
with $D=32/(3\pi)$, $A=3.2581$, $\beta=163.44$ and $\gamma=4.7125$. Notice that, with this definition, $0 \leq g_0(r_s) \leq 1/2$.
In Fig.~\ref{fig:limitmuinf}, we have plotted $\mu^2 \bar{\varepsilon}_{c}^{\mu}(r_s)$ computed with the CCD and FHNC methods with respect to $\mu$ for $r_s=2$.  This plot actually illustrates a general trend: for large values of $r_s$, the correlation energy per particle computed from the CCD method does not exhibit the correct behavior for $\mu \to \infty$. On the contrary, the FHNC method seems to perform better in this limit in spite of an important numerical noise. However, for small $r_s$ ($r_s \lesssim 1$), the CCD method becomes exact, since it reduces to the Random Phase Approximation (See Appendix~\ref{app:cc}), and thus respects the $\mu \to \infty$ limit.

The short-range correlation energies per particle for the \textit{erf} and \textit{erfgau} interactions are represented by the analytical parametrization
\begin{equation}
\label{fit}
\bar{\varepsilon}_{c}^{\mu}(r_s) = \frac{\varepsilon_{c}(r_s)}{1 + c_1(r_s)\mu + c_2(r_s)\mu^2},
\end{equation}
where $c_1(r_s)$ is determined by a fit
\begin{equation}
\label{}
c_1(r_s) = \frac{u_1 r_s + u_2 r_s^2}{1 + v_1 r_s},
\end{equation}
with $u_1=1.0271$, $u_2=-0.2302$, $v_1=0.6197$ for \textit{erf}, $u_1=0.3916$, $u_2=0.0223$, $v_1=0.9105$ for \textit{erfgau} using the CCD data and $u_1=0.4795$, $u_2=1.0094$, $v_1=10.1247$ for \textit{erfgau} using the FHNC data, and $c_2(r_s)$ is imposed by the exact limit for $\mu \to \infty$
\begin{equation}
\label{}
c_2(r_s) = \frac{8 r_s^3 \varepsilon_{c}(r_s)}{3 C (g_0(r_s)-1/2)}.
\end{equation}
These analytical parametrizations for \textit{erf} and \textit{erfgau} interactions are represented in Fig.~\ref{fig:ecsrerf} and~\ref{fig:ecsrerfgau} for $r_s=0.5$ and $2$. The two parametrizations for the \textit{erfgau} interaction differ only at small $r_s$ where the imposition of the exact $\mu \to \infty$ limit make both expressions close to the CCD data.

\begin{acknowledgments}
We are grateful to P. Gori-Giorgi (University ``La Sapienza'', Roma) for stimulating discussions.
\end{acknowledgments}

\appendix

\section{Exchange energy of the uniform electron gas with modified interaction}
\label{app:exchange}

The uniform electron gas can be considered as a system of $N$ electrons in a box of volume $\Omega$ with a uniform background of positive charge to ensure neutrality, studied in the thermodynamic limit (i.e $N \to \infty$ and $\Omega \to \infty$ such that the density $n=N/\Omega$ remains constant). This system is described by the electronic Hamiltonian
\begin{equation}
\label{}
\hat{H}=\hat{H}_0 + \hat{H}_{\text{int}},
\end{equation}
where $\hat{H}_0=\hat{T}$ is the kinetic energy operator and $\hat{H}_{\text{int}}$ is the electron-electron interaction which can be expressed by its Fourier expansion
\begin{equation}
\label{}
\hat{H}_{\text{int}} = \frac{1}{\Omega} \sum_{i<j} \sum_{\b{k} \not= \b{0}} v_{ee}(k) e^{i \b{k}.\b{r}_{ij}},
\end{equation}
where $v_{ee}(k)$ is the Fourier transform of the (modified) electron-electron interaction $v_{ee}(r)$. The term $\b{k} = \b{0}$ corresponding to the Hartree energy has been removed since it cancels out with the background energy and the electron-background interaction energy, provided that the same modified interaction has been applied to all these terms. The exchange energy corresponds to the first-order correction
\begin{equation}
\label{}
E_{x}= \bra{\Phi} \hat{H}_{\text{int}} \ket{\Phi},
\end{equation}
where $\Phi$ is the ground-state wave function of Hamiltonian $\hat{H}_0$ (a Slater determinant of plane-waves). It has been shown~\cite{FetWal-BOOK-03} that the exchange energy reduces to
\begin{equation}
\label{Ex}
E_{x}= - \frac{1}{12 \pi^4} k_{F}^3 \int_{0}^{\infty} q^2 v_{ee}(q) \left( 1 -\frac{3}{2} x + \frac{1}{2} x^3 \right) \theta(1-x) d q ,
\end{equation}
with $k_F = (3 \pi^2 n)^{1/3}$, $x=q/(2 k_F)$, $\theta(y)=1$ if $y>0$ and $\theta(y)=0$ if $y<0$. 

The Fourier transform of the Coulomb interaction is
\begin{equation}
\label{}
v_{ee,\coul}(q)=\frac{4\pi}{q^2},
\end{equation}
so that Eq.~(\ref{Ex}) leads after integration over $x$ the well known exchange energy per particle $\varepsilon_x=E_x/N$
\begin{equation}
\label{}
\varepsilon_{x,\coul}(r_s)=-\frac{3}{8} \left( \frac{18}{\pi^2} \right)^{1/3} \frac{1}{r_s},
\end{equation}
with $r_s=1/(\alpha k_F)$ and $\alpha=(4/(9\pi))^{1/3}$. For the \textit{erf} interaction, inserting the Fourier transform
\begin{equation}
\label{}
v_{ee,\erf}^{\mu}(q)=\frac{4\pi}{q^2} e^{- q^2/(4 \mu^2)}
\end{equation}
into Eq.~(\ref{Ex}) leads to the long-range exchange energy per particle
\begin{widetext}
\begin{equation}
\label{}
\varepsilon_{x,\erf}^{\mu}(r_s)=- \left( \frac{18}{\pi^2} \right)^{1/3} \frac{1}{r_s}
A \left( \sqrt{\pi} \erf \left( \frac{1}{2A} \right) +(2A-4A^3) e^{-1/(4A^2)} -3A + 4A^3 \right),
\end{equation}
where $A=\mu/(2k_F)$. The short-range exchange energy per particle is then
\begin{eqnarray}
\bar{\varepsilon}_{x,\erf}^{\mu}(r_s) &=& \varepsilon_{x}(n) - \varepsilon_{x,\erf}^{\mu}(n)
\nonumber\\
&=& - \left( \frac{18}{\pi^2} \right)^{1/3} \frac{1}{r_s}
\Biggl[ \frac{3}{8} -
A \left( \sqrt{\pi} \erf{\frac{1}{2A}} +(2A-4A^3) e^{-1/(4A^2)} -3A + 4A^3 \right)
\Biggl].
\label{exsrerf}
\end{eqnarray}
\end{widetext}
Similarly, the Fourier transform of the \textit{erfgau} interaction writes
\begin{equation}
\label{}
v_{ee,\erfgau}^{\mu}(q)=\frac{4\pi}{q^2} e^{- q^2/(4 \mu^2)} -\frac{6\sqrt{3}\pi}{\mu^2} e^{- 3 q^2/(4 \mu^2)},
\end{equation}
so that the long-range exchange energy per particle is
\begin{widetext}
\begin{equation}
\begin{split}
\label{}
\varepsilon_{x,\erfgau}^{\mu}(r_s)=- \left( \frac{18}{\pi^2} \right)^{1/3} \frac{1}{r_s}
\Biggl[
A \left( \sqrt{\pi} \erf \left( \frac{1}{2A} \right) +(2A-4A^3) e^{-1/(4A^2)} -3A + 4A^3 \right)  \\
- A
\left( \sqrt{\pi} \erf \left( \frac{1}{2B} \right) +(2B-16B^3) e^{-1/(4B^2)} -6B +16B^3 \right)
\Biggl],
\end{split}
\end{equation}
and the short-range exchange energy per particle is
\begin{equation}
\begin{split}
\label{exsrerfgau}
\bar{\varepsilon}_{x,\erfgau}^{\mu}(r_s)=- \left( \frac{18}{\pi^2} \right)^{1/3} \frac{1}{r_s}
\Biggl[
 \frac{3}{8} -
A \left( \sqrt{\pi} \erf \left( \frac{1}{2A} \right) +(2A-4A^3) e^{-1/(4A^2)} -3A + 4A^3 \right) \\
+ A
\left( \sqrt{\pi} \erf \left( \frac{1}{2B} \right) +(2B-16B^3) e^{-1/(4B^2)} -6B +16B^3 \right)
\Biggl],
\end{split}
\end{equation}
\end{widetext}
where $B=\mu/(2 \sqrt{3} k_F)$.

\section{Coupled-Cluster calculations of the uniform electron gas with modified interaction}
\label{app:cc}
For the Coulomb interaction, Freeman~\cite{Fre-PRB-77} has calculated the correlation energy of the uniform electron gas by summing the ring and screened exchange diagrams using the Coupled-Cluster method with double excitations (CCD). In this appendix, we rapidly give the corresponding equations for an arbitrary electron-electron interaction $v_{ee}(r)$.

The CCD wave function is constructed from the non-interacting determinant of plane waves $\Phi$ through
\begin{equation}
\label{PsiCCD}
\ket{\Psi} = e^{\hat{T_2}} \ket{\Phi},
\end{equation}
where the excitation operator $\hat{T_2}$ is expressed in second quantization notation as
\begin{equation}
\label{}
\hat{T_2} = \sum_{\b{k}_i,\b{k}_j,\b{q}} t_{\b{q}}(\b{k}_i,\b{k}_j) a^{\dag}_{\b{k}_i+\b{q}} a^{\dag}_{\b{k}_j-\b{q}} a_{\b{k}_j} a_{\b{k}_i}.
\end{equation}
Retaining only the ring diagrams, the amplitudes $t_{\b{q}}(\b{k}_i,\b{k}_j)$ are solutions of the equations (with momentum in $k_F$ units)
\begin{widetext}
\begin{eqnarray}
t_{\b{q}}(\b{k}_i,\b{k}_j) &=& \frac{v_{ee}(q)}{3 \pi^2 k_F D_{\b{q}}(\b{k}_i,\b{k}_j)}
\Biggl[
1 + 6 \pi^2 \int \frac{d\b{k}}{(2\pi)^3} 
(t_{\b{q}}(\b{k}_i,\b{k}) + t_{\b{q}}(\b{k}_j,\b{k})) \theta(1-k) \theta(|\b{k}+\b{q}|-1)
\nonumber\\
&&+ 18 \pi^4 \int \frac{d\b{k}}{(2\pi)^3} \int \frac{d\b{k'}}{(2\pi)^3}
(t_{\b{q}}(\b{k}_i,\b{k}) t_{\b{q}}(\b{k}_j,\b{k'}) + t_{\b{q}}(\b{k}_i,\b{k'}) t_{\b{q}}(\b{k}_j,\b{k})) 
\nonumber\\
&&\times \theta(1-k) \theta(1-k') \theta(|\b{k}+\b{q}|-1) \theta(|\b{k'}+\b{q}|-1)
\Biggl],
\label{tqequation}
\end{eqnarray}
\end{widetext}
with $D_{\b{q}}(\b{k}_i,\b{k}_j) = -(q^2 + \b{q}.(\b{k}_i + \b{k}_j))$. Compared to the original work of Freeman, the only modification appears in the Fourier transform $v_{ee}(q)$ of the arbitrary electron-electron interaction $v_{ee}(r)$.
Once the amplitudes have been computed, the correlation energy per particle can be calculated by
\begin{equation}
\label{}
\varepsilon_{c} = \varepsilon_{c,dir} + \varepsilon_{c,ex},
\end{equation}
where $\varepsilon_{c,dir}$ and $\varepsilon_{c,ex}$ are the direct and exchange contributions given by
\begin{widetext}
\begin{eqnarray}
\varepsilon_{c,dir} &=& 18 \pi^4 k_F \int \frac{d\b{q}}{(2\pi)^3} 
\int \frac{d\b{k}_i}{(2\pi)^3} \int \frac{d\b{k}_j}{(2\pi)^3}
v_{ee}(q) t_{\b{q}}(\b{k}_i,\b{k}_j)
\nonumber\\
&&\times \theta(1-k_i) \theta(1-k_j) \theta(|\b{k}_i+\b{q}|-1) \theta(|\b{k}_j+\b{q}|-1),
\label{}
\end{eqnarray}
\begin{eqnarray}
\varepsilon_{c,ex} &=& -9 \pi^4 k_F \int \frac{d\b{q}}{(2\pi)^3}
\int \frac{d\b{k}_i}{(2\pi)^3} \int \frac{d\b{k}_j}{(2\pi)^3}
v_{ee}(|\b{k}_i + \b{k}_j + \b{q}|) t_{\b{q}}(\b{k}_i,\b{k}_j)
\nonumber\\
&&\times \theta(1-k_i) \theta(1-k_j) \theta(|\b{k}_i+\b{q}|-1) \theta(|\b{k}_j+\b{q}|-1).
\label{}
\end{eqnarray}
\end{widetext}
The direct contribution, corresponding to the ring diagrams, is the usual correlation energy within the Random Phase Approximation (RPA). The exchange contribution includes additional screened exchange diagrams. In the high-density limit ($r_s \to 0$), the method reduces to the RPA and thus becomes exact.

In practice, it is convenient to introduce the intermediate quantity
\begin{equation}
\label{}
T_{q}(\b{k}_i) = \int \frac{d\b{k}}{(2\pi)^3} t_q(\b{k}_i,\b{k}) \theta(1-k) \theta(|\b{k}+\b{q}|-1),
\end{equation}
and to perform the integration by Gauss-Legendre quadrature. Equation~(\ref{tqequation}) is then equivalent to 
\begin{equation}
\label{Tequation}
\sum_{j} A_{ij} T_j = B_i,
\end{equation}
with $T_j = T_{q}(\b{k}_j)$ and
\begin{equation}
\label{}
A_{ij} = \delta_{ij} \left( 1 -\frac{2}{k_F} \sum_{m} W_m D_{im} \right) -\frac{2}{k_F} W_j D_{ij},
\end{equation}
\begin{equation}
\label{}
B_i = \sum_{m} W_m \left(  \frac{D_{im}}{3\pi^2 k_F} +\frac{12\pi^2}{k_F} T_i T_m D_{im}\right),
\end{equation}
where $D_{ij} = v_{ee}(q)/D_q(\b{k}_i,\b{k}_j)$ and $W_m$ are the quadrature weights. As $B_i$ actually depends on the $T_i$'s coefficients, Eq.~(\ref{Tequation}) have to be solved iteratively.

\section{Fermi-hypernetted chain theory for homogeneous systems}
\label{app:fhnc}

Similar in spirit to the CCD approach~(\ref{PsiCCD}), the Fermi-hypernetted chain (FHNC) method \cite{Cla-INC-79} is based on an approximate product ansatz for the wave function
\begin{equation}
 \Psi \left( {\bf r}_1, {\bf r}_2 , \ldots ,{\bf r}_N \right)
 = \exp \left[ \sum_{i<j} u_2({\bf r}_i, {\bf r}_j) \right]
 \Phi \left( {\bf r}_1 , {\bf r}_2 , \ldots ,{\bf r}_N \right) ,
\label{prod}
\end{equation}
where the correlation factor, called Jastrow factor, acts on a single Slater determinant $\Phi$. For homogeneous systems the pair-correlation function $u_2$ depends only on the inter-electron coordinate $r_{ij}$.
The close relationship between CCD and FHNC methods is not restricted to a purely formal analogy between the pair-correlation function $u_2$ and the CCD excitation operator $\hat{T_2}$. This topic has been extensively discussed in a review article by Bishop~\cite{Bis-TCA-91}.
For bosonic systems, both methods are actually equivalent on a certain level of approximation.

It is an important feature of the Jastrow ansatz that the exact short- and long-range asymptotic behavior
of a homogeneous system can be expressed as simple functions of the inter-electron coordinate.
In the case of a Coulomb potential, Kato's cusp condition for electrons with antiparallel spin 
imposes a constraint on the first derivative of the pair-correlation function 
\begin{equation}
 \left. \frac{d u_2(r_{12})}{d r_{12}} \right|_{r_{12}=0} = \frac{1}{2} ,
\label{Kato}
\end{equation}
which can be exactly represented by a Jastrow factor.
We discuss below how the modified interaction affects the short-range behavior of the Jastrow factor.
The long-range asymptotic behavior of electron correlations is well described by the RPA approximation \cite{Ful-BOOK-93}.
It provides an explicit asymptotic expression for the pair-correlation function
\begin{equation}
 \lim_{r_{12} \rightarrow \infty} \; u_2(r_{12}) = - \frac{1}{\omega_{\mbox{\small pl}} \, r_{12}} ,
\label{plasmon}
\end{equation}
where the plasmon frequency $\omega_{\mbox{\small pl}} = \sqrt{4 \pi n}$ of the electron gas enters into the denominator.
This asymptotic behavior can be reproduced by the FHNC method~\cite{Kro-PRA-77}.

For a given pair-correlation function, the FHNC equations represent a nonlinear system of equations between ``nodal'' $N(r_{12})$, 
``non-nodal'' $X(r_{12})$ and ``elementary'' $E(r_{12})$ functions.
Each of these functions can be expressed as an infinite sum of certain types of diagrams build up from 
the pair-correlation function and the one-particle density matrix of the noninteracting system. 
Some of these equations are conveniently expressed in coordinate 
space, the others in momentum space. The system of equations is underdetermined and requires an a priori knowledge of the
``elementary'' diagrams in order to get a unique solution. 
In a series of papers Krotscheck developed a consistent approximation scheme for the FHNC equations 
\cite{Kro-JLTP-77,Kro-AP-84,KroKohQia-PRB-85},
which preserves the correct asymptotic behavior on each level of approximation.
We have used the FHNC//0 method which corresponds to the lowest level of approximation, where
``elementary'' diagrams are neglected altogether. The FHNC//0 equations are given by 
\begin{eqnarray}
 \Gamma_{dd}(r_{12}) &:=& X_{dd}(r_{12}) + N_{dd}(r_{12}) 
\nonumber\\
&=& \exp \left[ 2 \, u_2(r_{12}) + N_{dd}(r_{12}) \right] -1 ,
\label{FHNC1}
\end{eqnarray}
\begin{equation}
 \tilde{N}_{dd}(k) = \tilde{X}_{dd}(k) \, S_F(k) \, \tilde{\Gamma}_{dd}(k) ,
\label{FHNC2}
\end{equation}
where $S_F$ is the liquid structure function of the noninteracting system.
We have used the dimensionless Fourier transform
\begin{equation}
 \tilde{f}(k) = n \int \! d \b{r} \, f(r) \, \exp(i {\bf k} \cdot {\bf r}).
\end{equation}
The ``nodal'' and ``non-nodal'' functions provide a link between the Jastrow factor and the
liquid structure function of the interacting system
\begin{equation}
 S(k) = S_F(k) + S_F(k)^2 \tilde{\Gamma}_{dd}(k) ,
\label{S}
\end{equation}
which is essentially the Fourier transform of the pair-density $n_2(r_{12})$
\begin{equation}
 S(k) = 1 + n \int \! d \b{r}_{12} \, \left( n_2(r_{12})/n^2 - 1 \right) \, \exp(i {\bf k} \cdot {\bf r}_{12}) .
\label{rho2}
\end{equation}
This connection enables an approximate variational treatment of the Jastrow factor within FHNC theory.

In the following we want to give a brief outline of the FHNC//0 optimization cycles
following essentially Krotscheck's paper \cite{KroKohQia-PRB-85}.
Starting point is an effective particle-hole potential
\begin{eqnarray}
 V_{ph} (r_{12}) &=& \left[ 1+ \Gamma_{dd} (r_{12}) \right] \, v_{ee} (r_{12}) 
+ \left| \nabla \left[ 1+ \Gamma_{dd} (r_{12}) \right]^{1/2} \right|^2
\nonumber\\
& &
 + \Gamma_{dd} (r_{12}) \, \omega_I (r_{12}) ,
\label{Vph}
\end{eqnarray}
which depends, beside diagrammatic contributions, on the bare (modified) interaction potential $v_{ee}$ and an 
induced interaction $\omega_I$. In momentum space the induced interaction
\begin{equation}
 \tilde{\omega}_I (k) = -\frac{k^2}{4} \left[ 1 + 2 \frac{S(k)}{S_F(k)} \right]
 \left[ \frac{1}{S(k)} - \frac{1}{S_F(k)} \right]^2 ,
\label{wI}
\end{equation}
can be expressed in terms of the liquid structure functions of the interacting and noninteracting system.
Within the high density regime, $v_{ee}$ can be taken as an initial guess for $V_{ph}$. 
Performing FHNC calculations at successively lower densities it is possible to reach the low density regime
by taking $V_{ph}$ from a slightly higher density as an initial guess in the optimization process. 
The particle-hole potential is related to the liquid structure function 
\begin{equation}
 S(k) = \frac{S_F(k)}{\left[ 1+ (4/k^2) \, S^2_F(k) \, \tilde{V}_{ph}(k) \right]^{1/2}} .
\label{Sk}
\end{equation}
In the first step of the optimization cycle Eq.~(\ref{Sk}) is used to get an improved approximation of 
the liquid structure function. Using Eqs.~(\ref{S}) and (\ref{wI}) it is now possible to obtain 
improved approximations for the induced interaction $\omega_I$ and the diagrammatic quantity $\Gamma_{dd} (r_{12})$.
These can be used in the second step to calculate an improved approximation of the 
particle-hole potential $V_{ph}$ (Eq.~\ref{Vph}).
The two steps provide a self-consistent optimization cycle, which can be repeated until convergence   
has been achieved. Finally we have used the FHNC Eqs.~(\ref{FHNC1}) and (\ref{FHNC2}) in order to obtain
the optimized FHNC//0 Jastrow factor.

\begin{figure}
\includegraphics[scale=0.45]{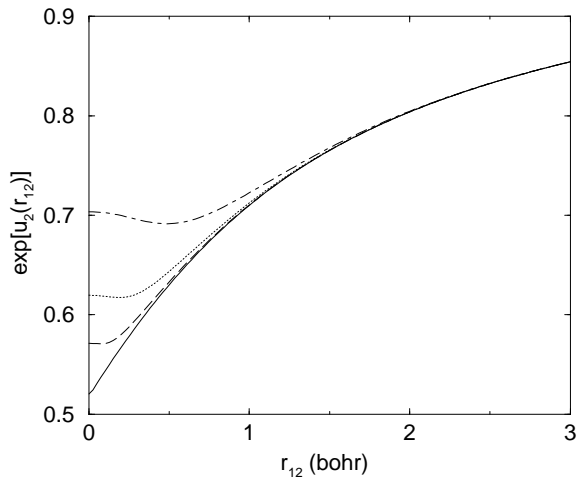}
\caption{Jastrow factors at $r_s=1$ for Coulomb interaction (solid line), and \textit{erfgau} interactions at interaction parameters
$\mu =20$ (dashed line), $\mu =10$ (dotted line) and $\mu =5$ (dotted-dashed line).
}
\label{fig:jastrow}
\end{figure}

Jastrow factors for the Coulomb interaction and the long-range \textit{erfgau} interaction are shown in Fig. \ref{fig:jastrow}. With decreasing value of the interaction strength $\mu$, the short-range part of the Jastrow factors is modified; it changes from a cusp at $r_{12} = 0$ for $\mu \to \infty$  into a smooth behavior for any finite $\mu$. For small values of $\mu$, a local minimum appears at an intermediate distance.
As expected, the long-range behavior of the Jastrow factor is not affected by the modified interaction.

\section{Divergence of calculations on the uniform electron gas with modified interaction}
\label{app:divergence}

\begin{figure}
\includegraphics[scale=0.5]{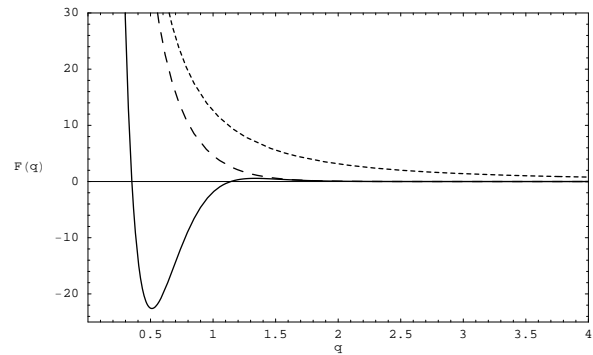}
\caption{Fourier transforms of the Coulomb interaction (dotted line), \textit{erf} interaction (dashed line) and \textit{erfgau} interaction (solid line), each plotted with a interaction parameter $\mu=0.5$.
}
\label{fig:fourier}
\end{figure}

With the \textit{erfgau} interaction, CCD and FHNC calculations of the uniform electron gas diverge for small values of $\mu$ and $r_s$. This is due to the particular form of the \textit{erfgau} interaction. In fact, whereas the Fourier transform of the Coulomb or \textit{erf} interaction is always positive, the Fourier transform of the \textit{erfgau} interaction
\begin{equation}
\label{}
v_{ee,\erfgau}(q)=\frac{4\pi}{q^2} e^{- q^2/(4 \mu^2)} -\frac{6\sqrt{3}\pi}{\mu^2} e^{- 3 q^2/(4 \mu^2)}
\end{equation}
can be negative (see Fig.~\ref{fig:fourier}). For small $\mu$, the negative part of $v_{ee,\erfgau}(q)$ is not negligible, introducing an attractive contribution to the electron-electron interaction. 

It is possible to estimate the domain of $\mu$ and $r_s$ for which CCD calculations do not converge because of this attractive interaction. Let's consider the ``state-average'' approximation to the CCD equations proposed by Bishop and L\"{u}hrmann~\cite{BisLuh-PRB-82}. This model consists in neglecting the exchange contribution to the correlation energy and averaging over the occupied momentum $\b{k}_i$ and $\b{k}_j$ the equations given in Appendix~\ref{app:cc}. The correlation energy per particle of the uniform electron gas is then written as (with momentum in $k_F$ units)
\begin{equation}
\label{}
\varepsilon_c =\frac{k_F}{4 \pi^2} \int_{0}^{\infty} dq q^2 v_{ee}(q) P(q) \langle t_q \rangle,
\end{equation}
where $P(q)=3q/4 - q^3/16$ if $q \leq 2$, $P(q)=1$ if $q > 2$, $v_{ee}(q)$ is the Fourier transform of the electron-electron interaction and $\langle t_q \rangle$ is the average of the amplitude $t_{\b{q}}(\b{k}_i,\b{k}_j)$ which is solution of the equation
\begin{equation}
\label{tqeq}
\langle t_q \rangle = \frac{1}{3 \pi^2 k_F} v_{ee}(q) 
\langle D_q^{-1} \rangle 
\left( 1+ \frac{\langle t_q \rangle}{P(q)} \right)^2,
\end{equation}
where $\langle D_q^{-1} \rangle$ is the average of the inverse of $D_{\b{q}}(\b{k}_i,\b{k}_j)$, introduced in Appendix~\ref{app:cc}.

\begin{figure}
\includegraphics[scale=0.7]{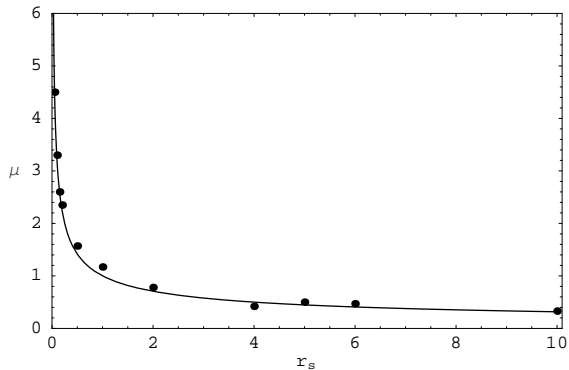}
\caption{Points at the limit of convergence for calculation of the correlation energy of the uniform electron gas with \textit{erfgau} interaction. The domain of divergence is well approximated by $\mu \sqrt{r_s} \lesssim 1$ (solid line).
}
\label{fig:divergencedomain}
\end{figure}

The general solution of~(\ref{tqeq}) is
\begin{equation}
\label{tq}
\langle t_q \rangle = \frac{ 1 - A(q) + \sqrt{1 -2 A(q)}}{A(q)P(q)},
\end{equation}
with $A(q)=(2 \langle D_q^{-1} \rangle  v_{ee}(q) P(q))/(3 \pi^2 k_F)$. Using the additional approximation $\langle D_q^{-1} \rangle \approx \langle D_q \rangle ^{-1} = -P(q)/q^2$, one sees immediately that this solution breaks down (more precisely, becomes imaginary) if
\begin{equation}
\label{ineq}
v_{ee}(q) < - \frac{3 \pi^2 k_F q^2}{4 P(q)^2}.
\end{equation}
Let's evaluate this inequality in the worst situation where $v_{ee}(q)$ and $P(q)$ take their minimum values. The interaction $v_{ee}(q)$ reaches its minimum $v_{min} \approx -5.6 k_F^2/\mu^2$ for $q \approx \mu/k_F$, and in this domain $P(q) \leq 3q/4$ so that condition~(\ref{ineq}) roughly gives
\begin{equation}
\label{divcond}
\mu \sqrt{r_s} \lesssim 1,
\end{equation}
where $r_s=1/(\alpha k_F)$ with $\alpha=(4/(9\pi))^{1/3}$ has been used.
In Fig.~\ref{fig:divergencedomain}, we have reported the values of $\mu$ and $r_s$ at the limit of convergence for the calculation of the correlation energy of the uniform electron gas with the \textit{erfgau} interaction, together with the divergence condition~(\ref{divcond}). Obviously, the domain of divergence is well approximated by this condition.

The divergence of both CCD and FHNC methods for these values of $\mu$ and $r_s$ where the attractive part of electron-electron interaction becomes important is reminiscent of the situation happening in a superconductor where the ordinary perturbation expansion breaks down for the superconducting phase.

\bibliographystyle{apsrev}
\bibliography{biblio}

\begin{thebibliography}{29}
\expandafter\ifx\csname natexlab\endcsname\relax\def\natexlab#1{#1}\fi
\expandafter\ifx\csname bibnamefont\endcsname\relax
  \def\bibnamefont#1{#1}\fi
\expandafter\ifx\csname bibfnamefont\endcsname\relax
  \def\bibfnamefont#1{#1}\fi
\expandafter\ifx\csname citenamefont\endcsname\relax
  \def\citenamefont#1{#1}\fi
\expandafter\ifx\csname url\endcsname\relax
  \def\url#1{\texttt{#1}}\fi
\expandafter\ifx\csname urlprefix\endcsname\relax\def\urlprefix{URL }\fi
\providecommand{\bibinfo}[2]{#2}
\providecommand{\eprint}[2][]{\url{#2}}

\bibitem[{\citenamefont{Hohenberg and Kohn}(1964)}]{HohKoh-PR-64}
\bibinfo{author}{\bibfnamefont{P.}~\bibnamefont{Hohenberg}} \bibnamefont{and}
  \bibinfo{author}{\bibfnamefont{W.}~\bibnamefont{Kohn}},
  \bibinfo{journal}{Phys. Rev.} \textbf{\bibinfo{volume}{{136}}},
  \bibinfo{pages}{B 864} (\bibinfo{year}{1964}).

\bibitem[{\citenamefont{Kohn and Sham}(1965)}]{KohSha-PR-65}
\bibinfo{author}{\bibfnamefont{W.}~\bibnamefont{Kohn}} \bibnamefont{and}
  \bibinfo{author}{\bibfnamefont{L.~J.} \bibnamefont{Sham}},
  \bibinfo{journal}{Phys. Rev. A} \textbf{\bibinfo{volume}{140}},
  \bibinfo{pages}{1133} (\bibinfo{year}{1965}).

\bibitem[{\citenamefont{Langreth and Perdew}(1977)}]{LanPer-PRB-77}
\bibinfo{author}{\bibfnamefont{D.~C.} \bibnamefont{Langreth}} \bibnamefont{and}
  \bibinfo{author}{\bibfnamefont{J.~P.} \bibnamefont{Perdew}},
  \bibinfo{journal}{Phys. Rev. B} \textbf{\bibinfo{volume}{{15}}},
  \bibinfo{pages}{2884} (\bibinfo{year}{1977}).

\bibitem[{\citenamefont{Langreth and Mehl}(1981)}]{LanMeh-PRL-81}
\bibinfo{author}{\bibfnamefont{D.~C.} \bibnamefont{Langreth}} \bibnamefont{and}
  \bibinfo{author}{\bibfnamefont{M.~J.} \bibnamefont{Mehl}},
  \bibinfo{journal}{Phys. Rev. Lett.} \textbf{\bibinfo{volume}{{47}}},
  \bibinfo{pages}{446} (\bibinfo{year}{1981}).

\bibitem[{\citenamefont{Langreth and Mehl}(1983)}]{LanMeh-PRB-83}
\bibinfo{author}{\bibfnamefont{D.~C.} \bibnamefont{Langreth}} \bibnamefont{and}
  \bibinfo{author}{\bibfnamefont{M.~J.} \bibnamefont{Mehl}},
  \bibinfo{journal}{Phys. Rev. B} \textbf{\bibinfo{volume}{{28}}},
  \bibinfo{pages}{1809} (\bibinfo{year}{1983}).

\bibitem[{\citenamefont{Perdew}(1985)}]{Per-PRL-85}
\bibinfo{author}{\bibfnamefont{J.~P.} \bibnamefont{Perdew}},
  \bibinfo{journal}{Phys. Rev. Lett.} \textbf{\bibinfo{volume}{{55}}},
  \bibinfo{pages}{1665} (\bibinfo{year}{1985}).

\bibitem[{\citenamefont{Perdew and Wang}(1986)}]{PerWan-PRB-86}
\bibinfo{author}{\bibfnamefont{J.~P.} \bibnamefont{Perdew}} \bibnamefont{and}
  \bibinfo{author}{\bibfnamefont{Y.}~\bibnamefont{Wang}},
  \bibinfo{journal}{Phys. Rev. B} \textbf{\bibinfo{volume}{{33}}},
  \bibinfo{pages}{8800} (\bibinfo{year}{1986}).

\bibitem[{\citenamefont{Perdew}(1986)}]{Per-PRB-86}
\bibinfo{author}{\bibfnamefont{J.~P.} \bibnamefont{Perdew}},
  \bibinfo{journal}{Phys. Rev. B} \textbf{\bibinfo{volume}{{33}}},
  \bibinfo{pages}{8822} (\bibinfo{year}{1986}).

\bibitem[{\citenamefont{Perdew}(1991)}]{Per-INC-91}
\bibinfo{author}{\bibfnamefont{J.~P.} \bibnamefont{Perdew}}, in
  \emph{\bibinfo{booktitle}{Electronic Structure of Solids '91}}, edited by
  \bibinfo{editor}{\bibfnamefont{P.}~\bibnamefont{Ziesche}} \bibnamefont{and}
  \bibinfo{editor}{\bibfnamefont{H.}~\bibnamefont{Eschrig}}
  (\bibinfo{publisher}{Akademie Verlag}, \bibinfo{address}{berlin},
  \bibinfo{year}{1991}).

\bibitem[{\citenamefont{Savin et~al.}(2003)\citenamefont{Savin, Colonna, and
  Pollet}}]{SavColPol-IJQC-03}
\bibinfo{author}{\bibfnamefont{A.}~\bibnamefont{Savin}},
  \bibinfo{author}{\bibfnamefont{F.}~\bibnamefont{Colonna}}, \bibnamefont{and}
  \bibinfo{author}{\bibfnamefont{R.}~\bibnamefont{Pollet}},
  \bibinfo{journal}{Int. J. Quantum. Chem.} \textbf{\bibinfo{volume}{{93}}},
  \bibinfo{pages}{166} (\bibinfo{year}{2003}).

\bibitem[{\citenamefont{Savin}(1996)}]{Sav-INC-96}
\bibinfo{author}{\bibfnamefont{A.}~\bibnamefont{Savin}}, in
  \emph{\bibinfo{booktitle}{Recent Developments of Modern Density Functional
  Theory}}, edited by \bibinfo{editor}{\bibfnamefont{J.~M.}
  \bibnamefont{Seminario}} (\bibinfo{publisher}{Elsevier},
  \bibinfo{address}{Amsterdam}, \bibinfo{year}{1996}), pp.
  \bibinfo{pages}{327--357}.

\bibitem[{\citenamefont{Leininger et~al.}(1997)\citenamefont{Leininger, Stoll,
  Werner, and Savin}}]{LeiStoWerSav-CPL-97}
\bibinfo{author}{\bibfnamefont{T.}~\bibnamefont{Leininger}},
  \bibinfo{author}{\bibfnamefont{H.}~\bibnamefont{Stoll}},
  \bibinfo{author}{\bibfnamefont{H.-J.} \bibnamefont{Werner}},
  \bibnamefont{and} \bibinfo{author}{\bibfnamefont{A.}~\bibnamefont{Savin}},
  \bibinfo{journal}{Chem. Phys. Lett.} \textbf{\bibinfo{volume}{{275}}},
  \bibinfo{pages}{151} (\bibinfo{year}{1997}).

\bibitem[{\citenamefont{Pollet et~al.}(2002)\citenamefont{Pollet, Savin,
  Leininger, and Stoll}}]{PolSavLeiSto-JCP-02}
\bibinfo{author}{\bibfnamefont{R.}~\bibnamefont{Pollet}},
  \bibinfo{author}{\bibfnamefont{A.}~\bibnamefont{Savin}},
  \bibinfo{author}{\bibfnamefont{T.}~\bibnamefont{Leininger}},
  \bibnamefont{and} \bibinfo{author}{\bibfnamefont{H.}~\bibnamefont{Stoll}},
  \bibinfo{journal}{J. Chem. Phys.} \textbf{\bibinfo{volume}{{4}}},
  \bibinfo{pages}{1250} (\bibinfo{year}{2002}).

\bibitem[{\citenamefont{Pollet et~al.}(2003)\citenamefont{Pollet, Colonna,
  Leininger, Stoll, Werner, and Savin}}]{PolColLeiStoWerSav-IJQC-03}
\bibinfo{author}{\bibfnamefont{R.}~\bibnamefont{Pollet}},
  \bibinfo{author}{\bibfnamefont{F.}~\bibnamefont{Colonna}},
  \bibinfo{author}{\bibfnamefont{T.}~\bibnamefont{Leininger}},
  \bibinfo{author}{\bibfnamefont{H.}~\bibnamefont{Stoll}},
  \bibinfo{author}{\bibfnamefont{H.-J.} \bibnamefont{Werner}},
  \bibnamefont{and} \bibinfo{author}{\bibfnamefont{A.}~\bibnamefont{Savin}},
  \bibinfo{journal}{Int. J. Quantum. Chem.} \textbf{\bibinfo{volume}{{91}}},
  \bibinfo{pages}{84} (\bibinfo{year}{2003}).

\bibitem[{\citenamefont{Stoll and Savin}(1985)}]{StoSav-INC-85}
\bibinfo{author}{\bibfnamefont{H.}~\bibnamefont{Stoll}} \bibnamefont{and}
  \bibinfo{author}{\bibfnamefont{A.}~\bibnamefont{Savin}}, in
  \emph{\bibinfo{booktitle}{Density Functional Method in Physics}}, edited by
  \bibinfo{editor}{\bibfnamefont{R.~M.} \bibnamefont{Dreizler}}
  \bibnamefont{and}
  \bibinfo{editor}{\bibfnamefont{J.}~\bibnamefont{da~Providencia}}
  (\bibinfo{publisher}{Plenum}, \bibinfo{address}{Amsterdam},
  \bibinfo{year}{1985}), pp. \bibinfo{pages}{177--207}.

\bibitem[{\citenamefont{Toulouse et~al.}()\citenamefont{Toulouse, Colonna, and
  Savin}}]{TouColSav-JJJ-XX}
\bibinfo{author}{\bibfnamefont{J.}~\bibnamefont{Toulouse}},
  \bibinfo{author}{\bibfnamefont{F.}~\bibnamefont{Colonna}}, \bibnamefont{and}
  \bibinfo{author}{\bibfnamefont{A.}~\bibnamefont{Savin}},
  \bibinfo{note}{submitted to Phys. Rev. A}.

\bibitem[{\citenamefont{Toulouse and Savin}()}]{TouSav-JJJ-XX}
\bibinfo{author}{\bibfnamefont{J.}~\bibnamefont{Toulouse}} \bibnamefont{and}
  \bibinfo{author}{\bibfnamefont{A.}~\bibnamefont{Savin}},
  \emph{\bibinfo{title}{in preparation}}.

\bibitem[{\citenamefont{Freeman}(1977)}]{Fre-PRB-77}
\bibinfo{author}{\bibfnamefont{D.~L.} \bibnamefont{Freeman}},
  \bibinfo{journal}{Phys. Rev. B} \textbf{\bibinfo{volume}{{15}}},
  \bibinfo{pages}{5512} (\bibinfo{year}{1977}).

\bibitem[{\citenamefont{Vosko et~al.}(1980)\citenamefont{Vosko, Wilk, and
  Nusair}}]{VosWilNus-CJP-80}
\bibinfo{author}{\bibfnamefont{S.~J.} \bibnamefont{Vosko}},
  \bibinfo{author}{\bibfnamefont{L.}~\bibnamefont{Wilk}}, \bibnamefont{and}
  \bibinfo{author}{\bibfnamefont{M.}~\bibnamefont{Nusair}},
  \bibinfo{journal}{Can. J. Phys.} \textbf{\bibinfo{volume}{{58}}},
  \bibinfo{pages}{1200} (\bibinfo{year}{1980}).

\bibitem[{\citenamefont{Burke et~al.}(1998)\citenamefont{Burke, Perdew, and
  Ernzerhof}}]{BurPerErn-JCP-98}
\bibinfo{author}{\bibfnamefont{K.}~\bibnamefont{Burke}},
  \bibinfo{author}{\bibfnamefont{J.~P.} \bibnamefont{Perdew}},
  \bibnamefont{and}
  \bibinfo{author}{\bibfnamefont{M.}~\bibnamefont{Ernzerhof}},
  \bibinfo{journal}{J. Chem. Phys} \textbf{\bibinfo{volume}{109}},
  \bibinfo{pages}{3760} (\bibinfo{year}{1998}).

\bibitem[{\citenamefont{Fetter and Walecka}(2003)}]{FetWal-BOOK-03}
\bibinfo{author}{\bibfnamefont{A.~L.} \bibnamefont{Fetter}} \bibnamefont{and}
  \bibinfo{author}{\bibfnamefont{J.~D.} \bibnamefont{Walecka}},
  \emph{\bibinfo{title}{Quantum Theory of Many-Particle Systems}}
  (\bibinfo{publisher}{Dover}, \bibinfo{year}{2003}).

\bibitem[{\citenamefont{Clark}(1979)}]{Cla-INC-79}
\bibinfo{author}{\bibfnamefont{J.}~\bibnamefont{Clark}}, in
  \emph{\bibinfo{booktitle}{Progress in Nuclear and Particle Physics Vol. 2}},
  edited by \bibinfo{editor}{\bibfnamefont{D.~H.} \bibnamefont{Wilkinson}}
  (\bibinfo{publisher}{Pergamon}, \bibinfo{address}{Oxford},
  \bibinfo{year}{1979}), p.~\bibinfo{pages}{89}.

\bibitem[{\citenamefont{Bishop}(1991)}]{Bis-TCA-91}
\bibinfo{author}{\bibfnamefont{R.}~\bibnamefont{Bishop}},
  \bibinfo{journal}{Theoret. Chim. Acta} \textbf{\bibinfo{volume}{{80}}},
  \bibinfo{pages}{95} (\bibinfo{year}{1991}).

\bibitem[{\citenamefont{Fulde}(1993)}]{Ful-BOOK-93}
\bibinfo{author}{\bibfnamefont{P.}~\bibnamefont{Fulde}},
  \emph{\bibinfo{title}{Electron Correlations in Molecules and Solids}}
  (\bibinfo{publisher}{Springer}, \bibinfo{address}{Berlin},
  \bibinfo{year}{1993}).

\bibitem[{\citenamefont{Krotscheck}(1977{\natexlab{a}})}]{Kro-PRA-77}
\bibinfo{author}{\bibfnamefont{E.}~\bibnamefont{Krotscheck}},
  \bibinfo{journal}{Phys. Rev. A} \textbf{\bibinfo{volume}{{15}}},
  \bibinfo{pages}{397} (\bibinfo{year}{1977}{\natexlab{a}}).

\bibitem[{\citenamefont{Krotscheck}(1977{\natexlab{b}})}]{Kro-JLTP-77}
\bibinfo{author}{\bibfnamefont{E.}~\bibnamefont{Krotscheck}},
  \bibinfo{journal}{J. Low Temp. Phys.} \textbf{\bibinfo{volume}{{27}}},
  \bibinfo{pages}{199} (\bibinfo{year}{1977}{\natexlab{b}}).

\bibitem[{\citenamefont{Krotscheck}(1984)}]{Kro-AP-84}
\bibinfo{author}{\bibfnamefont{E.}~\bibnamefont{Krotscheck}},
  \bibinfo{journal}{Ann. Phys.} \textbf{\bibinfo{volume}{{155}}},
  \bibinfo{pages}{1} (\bibinfo{year}{1984}).

\bibitem[{\citenamefont{Krotscheck et~al.}(1985)\citenamefont{Krotscheck, Kohn,
  and Qian}}]{KroKohQia-PRB-85}
\bibinfo{author}{\bibfnamefont{E.}~\bibnamefont{Krotscheck}},
  \bibinfo{author}{\bibfnamefont{W.}~\bibnamefont{Kohn}}, \bibnamefont{and}
  \bibinfo{author}{\bibfnamefont{G.-X.} \bibnamefont{Qian}},
  \bibinfo{journal}{Phys. Rev. B} \textbf{\bibinfo{volume}{{32}}},
  \bibinfo{pages}{5693} (\bibinfo{year}{1985}).

\bibitem[{\citenamefont{Bishop and L\"{u}hrmann}(1982)}]{BisLuh-PRB-82}
\bibinfo{author}{\bibfnamefont{R.~F.} \bibnamefont{Bishop}} \bibnamefont{and}
  \bibinfo{author}{\bibfnamefont{K.~H.} \bibnamefont{L\"{u}hrmann}},
  \bibinfo{journal}{Phys. Rev. B} \textbf{\bibinfo{volume}{{26}}},
  \bibinfo{pages}{5523} (\bibinfo{year}{1982}).

\end{thebibliography}

\end{document}